\documentclass[a4paper,oneside,11pt]{amsart}

\usepackage{amsmath,amssymb,epsf}
\usepackage{amsfonts,amsbsy,amscd,stmaryrd}
\usepackage{latexsym,amsopn}
\usepackage{amsthm}
\usepackage{enumitem}
\usepackage{mathabx}

\usepackage{avant}

\allowdisplaybreaks
 
\usepackage{graphicx}
\usepackage{color}
\definecolor{Blue}{rgb}{0.,0.,1.}
\definecolor{Red}{rgb}{1.,0.,0.}

\newcounter{smallarabics}
\newenvironment{arabicenumerate}
{\begin{list}{{\normalfont\textrm{(\arabic{smallarabics})}}}
  {\usecounter{smallarabics}\setlength{\itemindent}{0cm}
   \setlength{\leftmargin}{5ex}\setlength{\labelwidth}{4ex}
   \setlength{\topsep}{0.75\parsep}\setlength{\partopsep}{0ex}
   \setlength{\itemsep}{0ex}}}
{\end{list}}

\newcounter{smallroman}

\newenvironment{nouppercase}{%
  \renewcommand{\uppercasenonmath}[1]{}}{}

\newcommand{\ben}{\begin{arabicenumerate}}  
\newcommand{\een}{\end{arabicenumerate}}

\newtheorem{theorem}{Theorem}[section]

\newtheorem{proposition}[theorem]{Proposition}
\newtheorem{lemma}[theorem]{Lemma}
\newtheorem{definition}[theorem]{Definition}

\newtheorem{remark}[theorem]{Remark}
\newtheorem{example}[theorem]{Example}
\newcommand{\beq}{\begin{equation}}
\newcommand{\eeq}{\end{equation}}

\newcommand{\bea}{\begin{aligned}}
\newcommand{\eea}{\end{aligned}}

\newcommand{\bex}{\begin{example}}
\newcommand{\eex}{\end{example}}
\def\bel{\begin{lemma}}
\def\eel{\end{lemma}}
\def\bet{\begin{theoreme}}
\def\eet{\end{theoreme}}
\def\bed{\begin{definition}}
\def\eed{\end{definition}}
\def\ber{\begin{remark}}
\def\eer{\end{remark}}

\def\rr{{\mathbb R}}
\def\mm{{\mathbb M}}

\def\cc{{\mathbb C}}

\def\ss{{\mathbb S}}
\def\hh{{\mathbb H}}

\def\part{{\rm par}}

\def\bar{\overline}

\def\c0inf{C_0^\infty}

\def\proof{
\noindent{\bf Proof.}\ \ }

\usepackage{calrsfs}
\DeclareMathAlphabet{\pazocal}{OMS}{zplm}{m}{n}

\def\cS{{\pazocal S}}

\def\cD{{\pazocal D}}
\def\cU{{\pazocal U}}

\def\cM{{\pazocal M}}

\def\i{{\rm i}}

\let\Im\relax
\let\Re\relax

\DeclareMathOperator{\Im}{Im}
\DeclareMathOperator{\Re}{Re}

\newcommand{\qeds}{\qed\medskip}

\def \p{ \partial}
\def\12{\frac{1}{2}}
\def\14{\frac{1}{4}}

\def\e{{\rm e}}

\newcommand{\one}{\boldsymbol{1}}

\def\c{{\rm c}}

\def\12{\frac{1}{2}}

\def\e{{\rm e}}

\def\bep{\begin{proposition}}
\def\eep{\end{proposition}}

\makeatletter
\newcommand*{\defeq}{\mathrel{\rlap{%
                     \raisebox{0.3ex}{$\m@th\cdot$}}%
                     \raisebox{-0.3ex}{$\m@th\cdot$}}%
                     =}
                     
\makeatother
\makeatletter
\newcommand*{\eqdef}{=\mathrel{\rlap{%
                     \raisebox{0.3ex}{$\m@th\cdot$}}%
                     \raisebox{-0.3ex}{$\m@th\cdot$}}%
                     }
\makeatother
\newcommand{\traa}[1]{\mskip-6mu\upharpoonright_{#1}}

\def\cf{C^\infty}

\def\c{{\rm c}}

\def\dS{{\rm dS}}
\def\scri{\mathcal{I}}

\let\origmaketitle\maketitle
\def\maketitle{
  \begingroup
  \def\uppercasenonmath##1{} 
  \let\MakeUppercase\relax 
	\origmaketitle
  \endgroup
}

\newcommand{\mass}{{\alpha^2 + \nu^2}}
\newcommand{\mmass}{{\alpha^2 - \nu^2}}

\newcommand{\D}{\alpha}
\newcommand{\ex}{{\rm{\textit{e}}}}

\begin{document}

\title[Conformal extension of the Bunch-Davies state]{\Large Conformal extension of the Bunch-Davies state \\ across the de Sitter boundary}

\author{\large Micha{\l} \textsc{Wrochna}}
\address{Universit\'e Grenoble Alpes, Institut Fourier, UMR 5582 CNRS, CS 40700, 38058 Grenoble \textsc{Cedex} 09, France}
\email{michal.wrochna@univ-grenoble-alpes.fr}

\begin{abstract}In the setting of the massive Klein-Gordon equation on de Sitter space, we discuss Vasy's asymptotic data at conformal infinity in terms of plane waves. In particular, we derive a short-hand formula for reconstructing solutions from their asymptotic data. Furthermore, we show that the natural Hadamard state induced from future (or past) conformal infinity coincides with the Bunch-Davies state.
\end{abstract}

\begin{nouppercase}
\maketitle
\end{nouppercase}

\section{Introduction, main result}

\subsection*{Introduction} In Quantum Field Theory on curved spacetimes, a central problem is how to distinguish between positive and negative frequency solutions of the Klein-Gordon or Dirac equation on the level of asymptotic data.

The use of conformal methods, pioneered by Penrose \cite{penrose64,penrose65}, proved to be particularly successful in this respect. On asymptotically flat spacetimes, their mathematically rigorous implementation led to successive developments including the description of the symplectic space of solutions at null infinity \cite{ashtekar,DMP}, the interpretation of asymptotic data in terms of a characteristic Cauchy (or Goursat) problem \cite{friedlander,mason,HN}, and showing that the naturally arising decomposition into positive and negative frequencies defines a canonical, pure Hadamard state \cite{DMP,Mo1,Mo2,characteristic}. The techniques were generalized in various ways to other settings including Schwarzschild spacetime \cite{DMP2,nicolas}, and possibly massive Klein-Gordon fields on a class of cosmological spacetimes including the cosmological chart of de Sitter space \cite{DMP1}.

Recently also methods of scattering theory combined with time-dependent pseudo-differential calculus were used to obtain similar results for massive Klein-Gordon fields on asymptotically static spacetimes \cite{inout}.
 
An important case, however, to which none of the aforementioned methods directly applies are scalar fields on \emph{global} de Sitter space. While de Sitter space is naturally endowed with a conformal boundary $\scri$ with a past ($\scri_-$) and future ($\scri_+$) component, the asymptotic structure of solutions is manifestly different from that on, e.g., Minkowski space. 

A rigorous approach was proposed relatively recently by Vasy for the Klein-Gordon equation on a broad class of asymptotically de Sitter-like spacetimes. Let $d\geq 2$ be the spacetime's dimension, let $\alpha=\frac{d-1}{2}$, and let $\alpha^2+\nu^2$ be the squared mass, where $\nu$ is a parameter which in the present work is assumed to be real and non-zero. The basic observation is that after a conformal transformation, the Klein-Gordon operator smoothly extends across $\scri$ to an operator of mixed hyperbolic-elliptic type,  denoted in what follows by $P_\nu$. The main feature that follows from Vasy's analysis is that after multiplication by a suitable conformal factor, any smooth solution of the Klein-Gordon equation extends to a unique solution of $P_\nu u=0$ of the form
\beq\label{eq:theform}
(\mu+\i0)^{\i\nu} u^+ + (\mu-\i0)^{\i\nu} u^- + u^0
\eeq
near $\scri$, where $u^+$, $u^-$ and $u^0$ are smooth, and $\mu$ is a suitable boundary-defining function of $\scri=\{\mu=0\}$. The restrictions of $u^+$ and $u^-$ to either $\scri_+$ or $\scri_-$ determine the solution uniquely, and therefore can serve as `scattering data' (of course, with all the usual reservations towards using the term `scattering theory' on de Sitter-like spacetimes).

Distributions of the form \eqref{eq:theform} are singular at $\scri$ and the type of singularity, i.e.~$(\mu+\i0)^{\i\nu}$ vs. $(\mu-\i0)^{\i\nu}$, can be used to distinguish between positive and negative frequencies (note that the $u^0$ term plays no r\^ole in that respect). The general, rigorous statement following from \emph{propagation estimates near radial sets} is that if $u^-$ vanishes at $\scri_+$ (or at $\scri_-$) then it has a wave front set with only positive frequencies \cite{kerrds}, meant in the same way as in the microlocal formulation of the Hadamard condition. Furthermore, it was shown recently in \cite{VW} that the asymptotic data $u\mapsto(u^+,u^-)\traa{\scri_+}$ diagonalize the symplectic form of the classical field theory. All these ingredients were combined to construct Hadamard states from asymptotic data and also to \emph{conformally extend} two-point functions of Hadamard states across $\scri$, in the sense that they are first multiplied by conformal factors and then extended. 

\medskip

While these constructions are valid for a broad class of asymptotically de Sitter-like spacetimes, the purpose of the present paper is to investigate how precisely do they work in the special case of de Sitter space, for which more explicit methods are available. 

Let us first recall that on de Sitter space, there is a canonical way of giving sense to positive and negative frequency solutions (and hence, to particles and anti-particles) in terms of \emph{plane waves} $(x^+_\dS\cdot\xi)^{\i\nu-\D}$ and $(x^-_\dS\cdot\xi)^{\i\nu-\D}$, which play an analogous r\^ole to the plane waves on Minkowski space (see Subsect.~\ref{ss:planewaves} for the precise definition). On a rigorous level (for non-zero mass), this boils down to the existence of a distinguished Hadamard state called the \emph{Bunch-Davies state} (frequently also called the \emph{Bunch-Davies vacuum} or \emph{Euclidean vacuum}, it can be characterized as the unique \emph{Hartle-Hawking-Israel state} on de Sitter space). The link with plane waves is provided by the following remarkably elegant formula for its two-point function due to Bros and Moschella \cite{bros1}:
\beq\label{defbds}
\Lambda^+(x_\dS,y_\dS)={\rm const.} \int_{\scri_+}\omega(\xi) (x_\dS^-\cdot\xi)^{\i\nu-\D}(\xi\cdot y_\dS^+)^{-\i\nu-\D},
\eeq
where $x_\dS,y_\dS$ are points in de Sitter space and $\omega$ is a suitable form on $\scri_+$, see Subsect.~\ref{sec:integrals}. 

\subsection*{Main result} Our approach is based on the observation that the plane waves $(x^\pm_\dS\cdot\xi)^{\i\nu-\D}$ conformally extend to distributions denoted by $(x^\pm_\ss\cdot\xi)^{\i\nu-\D}$, which can be thought as plane waves for $P_\nu$. This allows us to conformally extend and compute the asymptotic data of various distributions expressed in terms of plane waves (see in particular Theorem \ref{thm:ext2} for the conformal extension of \eqref{defbds}), and to prove the following result. 

\begin{theorem}\label{thm:main} Assume $\nu\in\rr\setminus\{0\}$. In the case of $d$-dimensional de Sitter space, the following holds true.
\begin{itemize}[leftmargin=0.7cm]
\item[$(1)$] For any $(v^+,v^-)\in\cf(\scri_+)^2$, the solution $u$ of $(\Box\,+\mass) u=0$ with data $(v^+,v^-)$ at $\scri_+$ is given by 
\[
u(x_\dS)=\frac{1}{a(\nu)}\int_{\scri_+}\omega(\xi)\big(  (x_\dS^+\cdot \xi)^{\i\nu-\D} v^+(\xi) + (x_\dS^-\cdot \xi)^{\i\nu-\D} v^-(\xi) \big),
\]
where $a(\nu)=(2\pi)^\D 2^{-\i\nu}\Gamma(-\i\nu)/\Gamma(-\i\nu+\D)$. 
\item[$(2)$] The \emph{in} state and the \emph{out} state constructed in \cite{VW} both coincide with the Bunch-Davies state.
\end{itemize}
\end{theorem}

\noindent This shows that in the special case of de Sitter space, Vasy's asymptotic data does not only distinguish between positive and negative frequencies in a microlocal sense, but it also decouples positive and negative frequency plane waves. A further consequence is that Vasy's Feynman propagator (see \cite[Sec.~4]{positive}) coincides with the Feynman propagator canonically associated with the Bunch-Davies state. Most importantly, our computations provide a simple illustration of the constructions in \cite{poisson,VW}, which in the general asymptotically de-Sitter like case require the use of microlocal techniques.


\section{Conformal extension of solutions}

\subsection{Geometrical setup}
The starting point is to consider $1+d$-dimensional Minkowski space $\mm^{1,d}=\rr^{1+d}$ equipped with its canonical metric
\[
g_\mm=dx_0^2 - (d x_1^2 + \cdots +d x_d^2).
\]
We use the convention 
\[
x\cdot y = x_0 y_0 - x_1 y_1 - \cdots - x_d y_d
\]
for the Minkowskian product, and extend its definition to $\cc^{d+1}$ (as a bi-linear form), the complexification of $\mm^{1,d}$.

We denote by $\rho$ the Euclidean distance from the origin, i.e.
\[
\rho = (x_0^2+\cdots + x_d^2)^{\12}.
\]
We also consider the Minkowski distance function
\[
r= \left| x_0^2 -(x_1^2+\cdots +x_d^2)\right|^{\12}.
\]
For our purposes, $d$-dimensional de Sitter space $\dS^d$ (of radius $r>0$) is best defined as the hyperboloid 
\beq\label{hyp1}
\dS^d = \{ x\in \rr^{1+d} \ |  \ r(x)={\rm const}, \ x_{0}^2<x_1^2+\cdots + x_d^2 \},  
\eeq
equipped with the metric $g_\dS$ induced from $g_\mm$. On the other hand, hyperbolic space (of radius $r>0$) is either of the two  hyperboloids 
\beq\label{hyp2}
\hh_\pm^d = \{ x\in \rr^{1+d} \ |  \ r(x)={\rm const}, \ x_{0}^2>x_1^2+\cdots + x_d^2, \ \pm x_0>0 \},  
\eeq
equipped with the Riemannian metric defined as minus the metric induced from $g_\mm$. If not specified otherwise, we take the radius $r$ in \eqref{hyp1} and \eqref{hyp2} to be $1$. Whenever the embedding of hyperbolic space as a hypersurface in $\mm^{1,d}$ will not be needed or the distinction between $\hh_+^d$ and $\hh_-^d$ will be of no relevance, we will write $\hh^d$ instead of $\hh_\pm^d$.

We will use the notation $x=(x_0,\ldots,x_{d})$ for points in $\mm^{1,d}$, and we will write alternatively $x_\dS$, resp.~$x_\hh$ when $x\in \dS^d$, resp.~$x\in\hh^d$. For any $x\in \mm^{1,d}\setminus\{0\}$ we denote by $x_\ss$ the point on the unit sphere
\beq\label{eq:unitsphere}
\ss^d = \{ x\in \rr^{1+d} \ |  \ \rho(x)=1\}
\eeq
with the same angular coordinates as $x$ (i.e., $x_\ss$ is the point where the unit sphere intersects the half-line from the origin passing through $x$, see Fig.~1). 
\def\svgwidth{7cm}
\begin{figure}[h]\label{fig1}
\begingroup%
  \makeatletter%
  \providecommand\color[2][]{%
    \errmessage{(Inkscape) Color is used for the text in Inkscape, but the package 'color.sty' is not loaded}%
    \renewcommand\color[2][]{}%
  }%
  \providecommand\transparent[1]{%
    \errmessage{(Inkscape) Transparency is used (non-zero) for the text in Inkscape, but the package 'transparent.sty' is not loaded}%
    \renewcommand\transparent[1]{}%
  }%
  \providecommand\rotatebox[2]{#2}%
  \ifx\svgwidth\undefined%
    \setlength{\unitlength}{320.85bp}%
    \ifx\svgscale\undefined%
      \relax%
    \else%
      \setlength{\unitlength}{\unitlength * \real{\svgscale}}%
    \fi%
  \else%
    \setlength{\unitlength}{\svgwidth}%
  \fi%
  \global\let\svgwidth\undefined%
  \global\let\svgscale\undefined%
  \makeatother%
  \begin{picture}(1,1.00599969)%
    \put(0,0){\includegraphics[width=\unitlength]{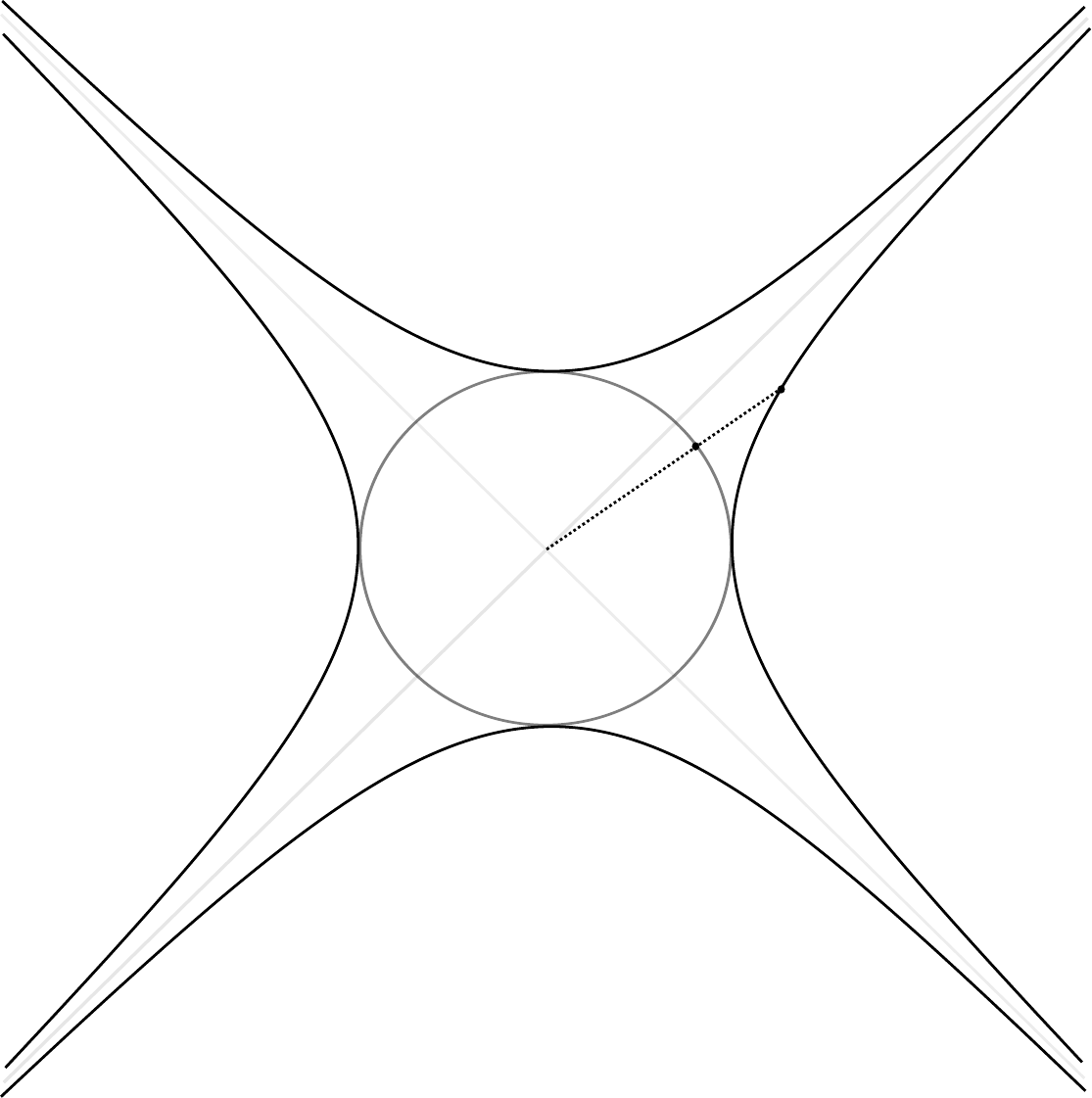}}%
    \put(0.30998022,0.75006664){\color[rgb]{0,0,0}\makebox(0,0)[lb]{\smash{$\hh_+^d$}}}%
    \put(0.31354215,0.23111462){\color[rgb]{0,0,0}\makebox(0,0)[lb]{\smash{$\hh_-^d$}}}%
    \put(0.21558804,0.3682502){\color[rgb]{0,0,0}\makebox(0,0)[lb]{\smash{$\dS^d$}}}%
    \put(0.73029229,0.35578347){\color[rgb]{0,0,0}\makebox(0,0)[lb]{\smash{$\dS^d$}}}%
    \put(0.71960641,0.6353978){\color[rgb]{0,0,0}\makebox(0,0)[lb]{\smash{$x$}}}%
    \put(0.6145284,0.61224503){\color[rgb]{0,0,0}\makebox(0,0)[lb]{\smash{$x_\ss$}}}%
  \end{picture}%
\endgroup%
\caption{The map $x\mapsto x_\ss\in\ss^d$, here for $x\in\dS^d$. Note that $\dS^d$ is connected since $d\geq 2$.}
\end{figure}
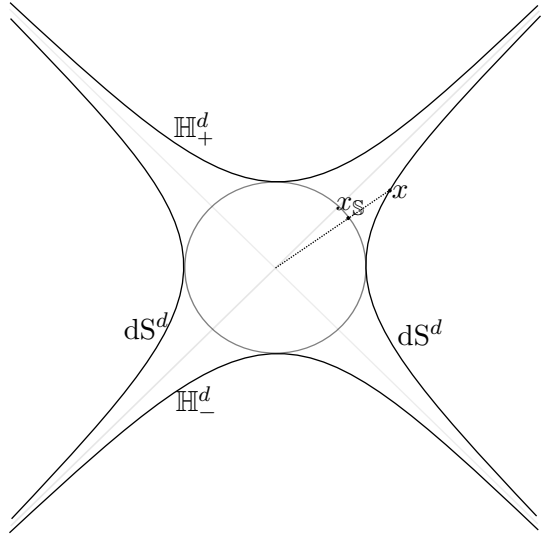
Using the map $x\mapsto x_\ss\in\ss^d$ we can identify the three hyperboloids $\hh_+^d$, $\dS^d$ and $\hh_-^d$ with three disjoint regions of the unit sphere $\ss^d$, namely with the intersection of \eqref{eq:unitsphere} with resp.~$\{x_0^2>x_1^2+\cdots+x_d^2\}\cap\{ x_0 > 0\}$, $\{x_0^2<x_1^2+\cdots+x_d^2\}$ and $\{x_0^2>x_1^2+\cdots+x_d^2\}\cap\{ x_0 < 0\}$. With these identifications, $\hh^d_+$, $\dS^d$, and $\hh^d_-$ cover the whole $\ss^d$ except for the two regions
\[
\scri_\pm\defeq \ss^d \cap \mathcal{C}_\pm^d, 
\]
(which are both  topologically $\ss^{d-1}$), where $\mathcal{C}_\pm^d$ is the future/past lightcone in $\mm^{1,d}$, i.e.
\[
\mathcal{C}_\pm^d=\mathcal{C}^d\cap\{ \pm x_0>0 \} , \  \ \mathcal{C}^d=\{ x\in \mm^{1,d} \ |  \ r(x)=0\}.
\]
It is therefore natural to define a manifold with boundary, denoted by $\bar\dS^d$, by endowing $\dS^d$ (identified with a region of $\ss^d$) with the boundary $\scri=\scri_+\cup \scri_-$. The connected component $\scri_+$, resp.~$\scri_-$, is called the \emph{future}, resp.~\emph{past conformal boundary} of $\bar\dS^d$, and their union $\scri$ is simply called the \emph{conformal boundary} (or \emph{conformal infinity}).  In an analogous way one also obtains a manifold with boundary $\bar\hh^d_\pm$ by considering $\hh^d_\pm$ as its interior, and $\scri_\pm$ as its boundary. 
\def\svgwidth{6cm}
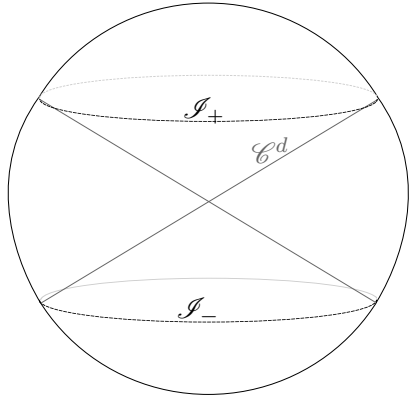
\begin{figure}[h]\label{fig2}
\begingroup%
  \makeatletter%
  \providecommand\color[2][]{%
    \errmessage{(Inkscape) Color is used for the text in Inkscape, but the package 'color.sty' is not loaded}%
    \renewcommand\color[2][]{}%
  }%
  \providecommand\transparent[1]{%
    \errmessage{(Inkscape) Transparency is used (non-zero) for the text in Inkscape, but the package 'transparent.sty' is not loaded}%
    \renewcommand\transparent[1]{}%
  }%
  \providecommand\rotatebox[2]{#2}%
  \ifx\svgwidth\undefined%
    \setlength{\unitlength}{456.00001221bp}%
    \ifx\svgscale\undefined%
      \relax%
    \else%
      \setlength{\unitlength}{\unitlength * \real{\svgscale}}%
    \fi%
  \else%
    \setlength{\unitlength}{\svgwidth}%
  \fi%
  \global\let\svgwidth\undefined%
  \global\let\svgscale\undefined%
  \makeatother%
  \begin{picture}(1,0.86661182)%
    \put(0,0){\includegraphics[width=\unitlength]{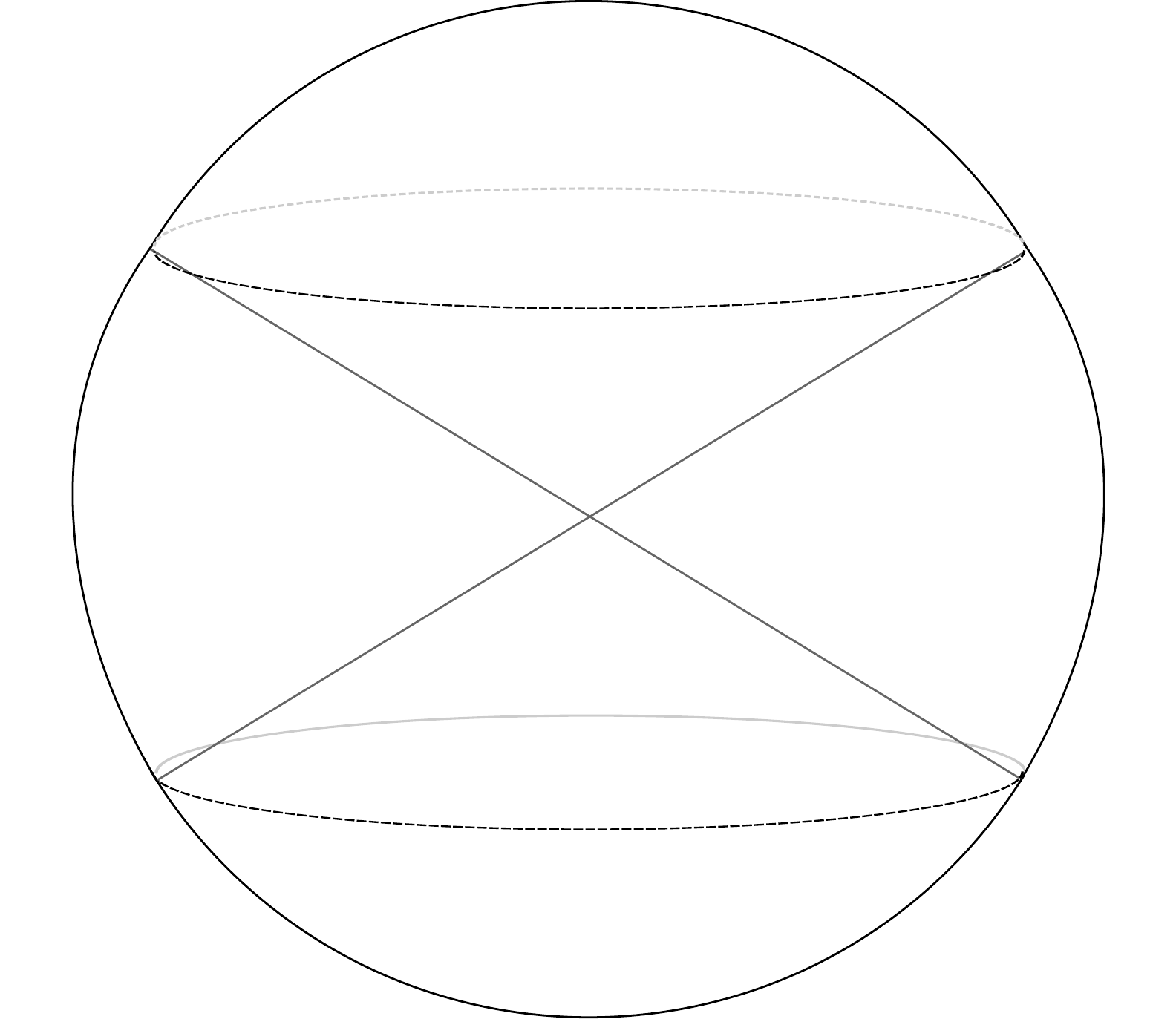}}%
    \put(0.44477831,0.61152685){\color[rgb]{0,0,0}\makebox(0,0)[lb]{\smash{$\scri_+$}}}%
    \put(0.43077883,0.17016066){\color[rgb]{0,0,0}\makebox(0,0)[lb]{\smash{$\scri_-$}}}%
    \put(0.59523805,0.51233455){\color[rgb]{0.4,0.4,0.4}\makebox(0,0)[lb]{\smash{$\mathcal{C}^d$}}}%
  \end{picture}%
\endgroup%
\caption{Extended de Sitter space. The `equatorial belt' part of the sphere $\ss^d$ (i.e., the region between $\scri_-$ and $\scri_+$) is identified with $\dS^d$, and the two `caps' resp.~below $\scri_-$ and above $\scri_+$ are identified with resp.~$\hh^d_-$ and $\hh^d_+$.}
\end{figure}
This way, 
\beq\label{eq:geom}
\ss^d = \bar{\hh}^d_+ \cup \bar\dS^d \cup \bar\hh^d_-, 
\eeq
where the boundary of $\bar{\hh}^d_\pm$ is identified with the $\scri_\pm$ component  of the boundary of $\bar\dS^d$ (see Fig.~2).

To get a more concrete description, let us set
\[
f(x)\defeq r \rho^{-1}(x), \ \ x_0^2 -(x_1^2+\cdots +x_d^2)\neq 0,
\]
and
\[
f_\dS\defeq f\traa{\dS^d}, \ \ f_\hh\defeq f\traa{\hh^d}.
\]
We stress that with our conventions, $r(x_\dS)=1$ and $r(x_{\hh})=1$. The stereographic projection $x\mapsto x_\ss$ identifies any $x_\dS\in\dS^d$ with the point $f(x_\dS)x_\dS$ on the unit sphere and any $x_\hh\in\hh^d$ with $f(x_\hh)x_\hh$. This justifies the short-hand notation
\[
x_\ss=f_\dS  x_\dS, \ \  x_\ss=f_\hh x_\hh,
\]
which will be used throughout the paper. 

The function $f_\dS$ is a convenient boundary-defining function of $\scri=\{f_\dS =0\}$ (understood as the boundary of $\dS^d$). Similarly, $f_\hh$ can be used as a boundary-defining function of $\scri_+=\{f_\hh =0\}$ (understood as the boundary of $\hh_+^d$) or  $\scri_-=\{f_\hh =0\}$ (understood as the boundary of $\hh_-^d$).


\subsection{Klein-Gordon equation and plane waves} \label{ss:planewaves}

We are primarily interested in solutions of the massive Klein-Gordon equation
\beq\label{eq:kg1}
\left(\Box_\dS + \mass \right)\phi=0,
\eeq
where $\Box_\dS$ is the Laplace-Beltrami (wave) operator on the de Sitter spacetime $(\dS^d,g_\dS)$,
\beq
\alpha=\frac{d-1}{2},
\eeq
and $\nu\in\rr$ will be assumed to be non-zero. The geometric considerations leading to \eqref{eq:geom} provide motivation for studying in parallel the equation
\beq\label{eq:akg2}
\left(-\Delta_{\hh_\pm} - \mmass\right)\phi=0,
\eeq
where $\Delta_{\hh_\pm}$ is the Laplace--Beltrami operator on $\hh^d_\pm$.

The equations \eqref{eq:kg1} and \eqref{eq:akg2} have special solutions that are named \emph{plane waves} because of various properties they share with the plane waves on Minkowski and Euclidean space. We will closely follow the exposition in \cite{bros2,MS,EM} (cf.~\cite{queva} for an introduction that discusses the zero-curvature limit and \cite{BJM} for a thorough exposition in the context of applications in interacting models in $1+1$ dimensions). First, for $\xi\in \mathcal{C}^d$ and $\lambda\in\cc$, let us introduce the notation
\[
(x^\pm\cdot\xi)^{\lambda}\defeq  \lim_{\mathcal{T}^\pm \ni z\to x}(z\cdot\xi)^{\lambda}
\]
where the boundary value is taken from the tuboid 
\[
\mathcal{T}^\pm=\{ x + \i y\in \cc^{d+1} | \ y\cdot y >0, \ \pm y_0 >0\}.
\]
The restriction of $(x^\pm\cdot\xi)^{\i\nu-\D}$ to $\dS^d$ yields a distribution which we denote by $(x_\dS^\pm\cdot\xi)^{\i\nu-\D}$, and similarly, the restriction of $(x^\pm\cdot\xi)^{\i\nu-\D}$ to either $\hh^d_+$ or $\hh^d_-$ is denoted by $(x_\hh^\pm\cdot\xi)^{\i\nu-\D}$. Note that if $\xi\in \mathcal{C}_\pm^d$, then $x_\hh\cdot\xi>0$ for $x_\hh\in\hh^d_\pm$, and thus on $\hh_\pm^d$ both $(x_\hh^+\cdot\xi)^{\i\nu-\D}$ and $(x_\hh^-\cdot\xi)^{\i\nu-\D}$ are simply the distribution $(x_\hh\cdot\xi)^{\i\nu-\D}$. The distributions $(x_\dS^\pm\cdot\xi)^{\i\nu-\D}$ and $(x_\hh\cdot\xi)^{\i\nu-\D}$ are called plane waves (and in the latter case are proportional to \emph{Eisenstein functions}, also called \emph{Eisenstein series}) and are solutions of
\[
\left(\Box_\dS + \mass \right)(x^\pm_\dS\cdot\xi)^{\i\nu-\D}=0, \ \ \left(\Delta_{\hh_\pm} + \mass\right)(x_\hh\cdot\xi)^{\i\nu-\D}=0,
\]
as can be shown by a simple computation, see e.g.~\cite{bros1,bros2,MS}.

In the context of the geometrical setup described in the previous subsection, it is also natural to introduce plane waves on $\ss^d$ as the restrictions 
\beq\label{eq:defpwss}
(x_\ss^\pm\cdot\xi)^{\i\nu-\D}=(x^\pm\cdot\xi)^{\i\nu-\D}\traa{\ss^d}.
\eeq
These new plane waves are related to their $\dS^d$ and $\hh^d$ counterparts by
\beq\label{relplw}
(x_\ss^\pm\cdot\xi)^{\i\nu-\D}=\begin{cases}f^{\i\nu-\D}_\dS (x^\pm_\dS\cdot\xi)^{\i\nu-\D} \ \mbox{on} \  \dS^d, \\ f^{\i\nu-\D}_{\hh} (x_\hh^\pm\cdot\xi)^{\i\nu-\D} \ \mbox{on} \  \hh^d, \end{cases}
\eeq
using the identification of $\dS^d$ and $\hh_+^d$ or $\hh_-^d$  with the respective regions of the sphere $\ss^d$. Therefore, we are allowed to think of \eqref{eq:defpwss} as extensions of the de Sitter plane waves  across the conformal boundary \emph{after multiplication by the conformal factor} $f^{\i\nu-\D}_{\dS}$. In what follows we will be interested in making this more systematic and extending more general classes of solutions in this sense.

First, let us point out the relation between the $\dS^d$/$\hh^d$/$\ss^d$ plane waves and the massless plane waves on $1+d$-dimensional Minkowski space
\[
\e^{\pm\i x \cdot \xi}=\e^{\pm\i \rho(x_\ss \cdot \xi)}=\begin{cases}\e^{\pm\i r(x_{\dS} \cdot \xi)} \ \mbox{on} \  \dS^d,\\ \e^{\pm\i r(x_{\hh} \cdot \xi)} \ \mbox{on} \  \hh^d.\end{cases}
\]
Namely, in terms of the Mellin transform $(\cM_r u)(\nu)\defeq \int_0^\infty r^{-\i\nu-1}u(r) d r$ in the Minkowski distance function $r$, one has
\beq\label{eq:mel1}
\bea
(x^\pm_{\dS}\cdot\xi)^{\i\nu-\D}&=\frac{1}{\Gamma(-\i\nu+\D)} \left(\cM_r  r^{\D} \e^{\pm\i r(x_{\dS} \cdot \xi)}\right)(\nu), \\
(x_{\hh}\cdot\xi)^{\i\nu-\D}&= \frac{1}{\Gamma(-\i\nu+\D)}\left(\cM_r  r^{\D} \e^{\pm\i r(x_{\hh} \cdot \xi)}\right)(\nu),
\eea
\eeq
in the sense of analytic continuation from $\Im\nu\gg0$. Unsurprisingly, an analogous relation is valid for $(x^\pm_\ss\cdot\xi)^{\i\nu-\D}$, i.e.
\beq\label{eq:mel2}
(x^\pm_\ss\cdot\xi)^{\i\nu-\D}=\frac{1}{\Gamma(-\i\nu+\D)}\left(\cM_\rho  \rho^{\D} \e^{\pm\i \rho(x_\ss \cdot \xi)}\right)(\nu).
\eeq
The identities \eqref{eq:mel1} are no coincidence. In fact, the wave operator $\Box_\mm$ on Minkowski space $\mm^{1,d}$ is related to the Klein-Gordon equation by
\[
\big(\cM_r r^{\D+2} \Box_\mm r^{-\D}\cM_r^{-1}\big)(\nu)=\begin{cases}\Box_\dS + \mass \ \mbox{on} \  \dS^d,\\ -\Delta_{\hh_\pm} - \mmass \ \mbox{on} \  \hh^d_\pm, \end{cases}
\]
see \cite{resolvent,poisson}. It turns out that if one now replaces $r$ by $\rho$, one obtains a differential operator on the sphere $\ss^d$ with \emph{smooth} coefficients (depending on the parameter $\nu$)
\[
P_{\nu}\defeq\big(\cM_\rho \rho^{\D+2} \Box_\mm \rho^{-\D}\cM_\rho^{-1}\big)(\nu),
\]
that satisfies\footnote{Note that with our conventions, $P_{\nu}$ is \emph{minus} the operator considered in \cite{poisson}.}
\beq\label{eq:whatsp}
P_{\nu}=\begin{cases}f^{\i\nu-\D-2}_\dS (\Box_\dS + \mass )f^{-\i\nu+\D}_\dS \ \mbox{on} \  \dS^d, \\ f^{\i\nu-\D-2}_{\hh} (-\Delta_{\hh_\pm} - \mmass)f^{-\i\nu+\D}_{\hh} \ \mbox{on} \  \hh_\pm^d, \end{cases}
\eeq
using again the identification of $\dS^d$ and $\hh_\pm^d$ with regions of $\ss^d$. This way $P_\nu$ is interpreted as the extension across the conformal infinity of the de Sitter Klein-Gordon operator, rescaled first with powers of the boundary defining function $f_\dS$ (needed to circumvent the divergence at the boundary).

By \eqref{relplw} and \eqref{eq:whatsp} the distributions $(x^\pm_\ss\cdot\xi)^{\i\nu-\D}$ satisfy $P_\nu (x^\pm_\ss\cdot\xi)^{\i\nu-\D}=0$, so one can think of these as plane waves for the conformally extended operator $P_\nu$. We stress that in view of \eqref{eq:whatsp}, while $P_\nu$ is hyperbolic in the de Sitter region, it is elliptic in the two copies of hyperbolic space. 

To see this explicitly on the level of the principal symbol of $P_\nu$, it is useful to introduce the coordinate 
\[
\mu\defeq \frac{x\cdot x}{\rho^2}=\frac{x_0^2 -(x_1^2+\cdots +x_d^2)}{x_0^2 +x_1^2+\cdots +x_d^2}.
\]
Then
\[
\mu=-f_{\dS}^2 \mbox{ on } \dS^d, \ \ \mu=f_{\hh}^2  \mbox{ on } \hh^d,
\]
so in particular $\mu<0$ on $\dS^d$ and $\mu>0$ on $\hh_+^d$ and $\hh_-^d$. In terms of the coordinate system obtained using $\mu$ and coordinates on $\ss^{d-1}$, the operator $P_\nu$ equals $-4\mu\p_\mu^2-\Delta_{\ss^{d-1}}$ plus a first order differential operator with smooth coefficients, where $\Delta_{\ss^{d-1}}$ is the Laplace--Beltrami operator on $\ss^{d-1}$, see \cite{resolvent} for details of the computation. 

The result which is central in our analysis (and which we will illustrate by means of wave packets) is a special case of a construction due to Vasy  \cite{kerrds,resolvent,poisson}, and states that any smooth solution of the de Sitter Klein-Gordon equation conformally extends across $\scri$ in a similar way as the plane waves \eqref{relplw}.

\begin{theorem}[\cite{poisson}]\label{thm:ext} Assume $\nu\in\rr\setminus\{0\}$. Suppose that $u\in\cf(\dS^d)$ solves 
\beq\label{eq:dflnj}
\left(\Box_\dS + \mass \right)u=0.
\eeq
Then $f^{\i\nu-\D}u$ extends to a unique distribution $\ex_\nu u$ on $\ss^d$ solving $P_\nu\ex_\nu u=0$.
\end{theorem}

\begin{definition}We will call $\ex_\nu u$ the \emph{conformal extension} of $u$.
\end{definition}

One of our goals will be to give a more explicit description of $\ex_\nu u$ when $u$ is a wave packet.

The dependence on the parameter $\nu$ is emphasized in the notation as a reminder that there are actually two different, but closely related ways of defining conformal extensions, namely, one can conformally extend a given solution of \eqref{eq:dflnj} either to a unique solution for $P_\nu$ (using the map $e_\nu$), or to a unique solution of $P_{-\nu}$ (using $e_{-\nu}$).

We stress that the conformal extension $\ex_\nu u$ of a non-zero smooth solution $u$ is \emph{not} smooth at $\scri=\{\mu=0\}$, despite $P_\nu$ having smooth coefficients there. Instead, it is shown in \cite{poisson} that it can be written near $\scri_+$ as
\beq\label{eq:expansion}
\ex_\nu u = u^0 + (\mu+\i 0)^{\i\nu}u^+ + (\mu-\i 0)^{\i\nu}u^-
\eeq
for some $u^0,u^+,u^-\in\cf(\ss^d)$. Furthermore, the pair of restrictions
\beq\label{eq:defrho}
\varrho_{\scri_+} u \defeq \begin{pmatrix} u^+\traa{\scri_+} \\ u^-\traa{\scri_+} \end{pmatrix}\in \cf(\scri_+)^2 
\eeq
determines $\ex_\nu u$ (and $u$) uniquely. We will often symbolically write statements such as \eqref{eq:expansion} in the form
\[
\ex_\nu u  \sim  v^0 + (\mu+\i 0)^{\i\nu}v^+ + (\mu-\i 0)^{\i\nu}v^-,
\]
where $v^0=u^0\traa{\scri_+}$, $v^\pm=u^\pm\traa{\scri_+}$.

These results are proved in \cite{poisson} in much greater generality, namely, they apply to conformally extended \emph{asymptotically} de Sitter spacetimes, provided that the metric satisfies an evenness condition at $\scri$ (stating that the conformally extended metric should be smooth in a coordinate which would correspond here to $\mu$ \cite{poisson}). In \cite{poisson} also complex values of $\nu$ are allowed, except for a discrete set of resonances.

 In the present context, the main merit of the whole construction is that the two components of the asymptotic data \eqref{eq:defrho} distinguish between positive and negative frequencies. While this can be made into a rigorous, general statement in the language of wave front sets and radial sets, our goal in what follows will be to explain what does this mean in terms of de Sitter plane waves.

\subsection{Integrals at $\scri$}\label{sec:integrals} Following \cite{bros1}, let us consider the differential form
\[
\omega= \sum_{i=1}^d (-1)^{i+1} \frac{\xi_i}{\xi_0}d\xi_1\wedge\cdots\wedge \widehat{d\xi_i}\wedge \cdots\wedge d\xi_d.
\]
An important result that we will use is \cite[Lem.~4.1]{bros1}, which states that if $v$ is a homogeneous function on $\mathcal{C}_+^d$ of degree $-(d-1)$, then $v\omega$ is a closed $(d-1)$-form on $\mathcal{C}_+^d$. In consequence, for any function $\psi$ on $\mathcal{C}_+^d$, homogeneous of degree $-\i\nu-\D$, we can form integrals such as
\beq\label{eq:integrals}
\int_{\scri_+}\omega(\xi)(x^\pm_\dS\cdot\xi)^{\i\nu-\D}\psi(\xi), \ \ \int_{\scri_+}\omega(\xi)(x^\pm_\ss\cdot\xi)^{\i\nu-\D}\psi(\xi),
\eeq
(recall that in our setting, $\scri_+$ is the intersection of $\mathcal{C}_+^d$ with the unit sphere in $\rr^{1+d}$) since $(x^\pm\cdot\xi)^{\i\nu-\D}\psi(\xi)$ is homogeneous of degree $-2\alpha=-(d-1)$ in $\xi$. A remarkable consequence of \cite[Lem.~4.1]{bros1} is that the integrals \eqref{eq:integrals} do not depend on the choice of ``orbital base'' of $\mathcal{C}_+^d$, meaning that we obtain the same result if we integrate over any smooth $(d-1)$-cycle $\gamma$ homotopic to $\scri_+$ in $\mathcal{C}_+^d$. Throughout the paper we will work with $\gamma=\scri_+$ only, in which case $\scri_+$ can be parametrized by $\ss^{d-1}\ni \hat{\xi} \mapsto (1,\hat\xi)/\sqrt{2}\in \mathcal{C}_+^d$, and then $\omega$ restricted to $\scri_+$ is proportional to the standard integration element of $\ss^{d-1}$.

As remarked in \cite[Rem. 2.2]{bros3}, if $\psi$ is smooth then the first integral \eqref{eq:integrals} defines a smooth function on $\dS^d$, and similarly the second one defines a smooth function on $\ss^d$. For this reason it is often more practical to work with smeared expressions such as \eqref{eq:integrals} rather than with plane waves directly.


\subsection{Asymptotics of plane waves} We will frequently use the formula
\beq\label{eq:mupm}
(\mu\pm\i 0)^{\i\nu}=\mu_+^{\i\nu} + \e^{\mp \nu\pi}\mu_-^{\i\nu},
\eeq
where $\mu_\pm^{\i\nu}=\theta(\pm \mu)|\mu|^{\i\nu}$ and $\theta$ is the Heaviside step function. Let us recall that $\mu<0$ on $\dS^d$ and $\mu>0$ on $\hh^d$, so the term with $\mu_-$ is relevant for the de Sitter region of $\ss^d$, whereas the $\mu_+$ term is relevant for $\hh^d$.

Let us complete $\mu$ with some coordinates in a chart neighborhood of $\scri_\pm$ in $\ss^d$. For $x\in\mm^{1,d}$, we define  
\[
\xi_\pm(x) =\lim_{\mu\to 0} x_{\ss} \in  \scri_\pm.
\]

Let $\xi\in \mathcal{C}_+^d$. We will show that the de Sitter and hyperbolic space plane waves have the asymptotics
\beq\label{ex:asdSH}
\bea
(x_\dS^\pm \cdot \xi)^{\i \nu-\D} \sim f^{-\i\nu+\D}_\dS (\xi_+(x)\cdot\xi)^{\i\nu - \D} &+ f^{\i\nu+\D}_\dS \e^{\mp \nu \pi} a(\nu) \delta(\xi_+(x),\xi),\\
(x_\hh \cdot \xi)^{\i \nu-\D} \sim f^{-\i\nu+\D}_\hh (\xi_+(x)\cdot\xi)^{\i\nu - \D} &+  f^{\i\nu+\D}_\hh a(\nu) \delta(\xi_+(x),\xi).
\eea
\eeq
at $\scri_+$, where
\beq
a(\nu)=\frac{(2\pi)^\D 2^{-\i\nu}\Gamma(-\i\nu)}{\Gamma(-\i\nu+\D)}.
\eeq
The asymptotics \eqref{ex:asdSH} are understood in the sense that for any $\psi\in\cf(\scri_+)$, homogeneous of degree $-\i\nu-\D$, there exists $u^\pm_\psi\in\cf(\dS^d)$ such that
\[
\int_{\scri_+}\omega(\xi)(x_\dS^\pm \cdot \xi)^{\i \nu-\D}\psi(\xi) = f^{-\i\nu+\D} u^-_\psi(x_\dS) + f^{\i\nu+\D} u^+_\psi(x_\dS),
\]
and
\[
u^-_\psi \traa{\scri_+}= \int_{\scri_+} \omega(\xi)(\xi_+(x)\cdot\xi)^{\i\nu - \D}\psi(\xi), \ \ u^+_\psi \traa{\scri_+}=\e^{\mp \nu \pi} a(\nu)\psi(\xi_+(x)),
\]
and similarly for the hyperbolic space version. A variant of the second statement in \eqref{ex:asdSH} was shown in \cite[Sec.~4.4]{MS} for a different choice of boundary-defining function of $\scri_+$ (cf.~\cite{BB} for a recent analysis of asymptotic properties of de Sitter wave packets).

Since $f^2=-\mu$ on $\dS^d$ and $f^2=\mu$ on $\hh^d$, using \eqref{eq:mupm} we can deduce the asymptotics \eqref{ex:asdSH} from the following `two-sided' result.
\begin{lemma} At  $\scri_+$ we have the asymptotics
\beq\label{eq:asss1}
(x_\ss^\pm \cdot \xi)^{\i \nu-\D} \sim (\xi_+(x)\cdot\xi)^{\i\nu - \D} + (\mu(x)\pm\i 0)^{\i\nu} a(\nu) \delta(\xi_+(x),\xi).
\eeq
\end{lemma}
\proof Let $c>0$. By \cite[Sec.~2.8]{gelfand}, the inverse Fourier transform of the distribution
\[
\rr^{2\alpha}\ni y \mapsto (c^2 + |y|^2)^{\i\nu-\D}
\]
is the distribution
\[
\rr^{2\alpha}\ni \eta \mapsto \frac{2^{\i\nu-\D+1}}{\Gamma(-\i\nu+\D)}\left(\frac{c}{|\eta|}\right)^{\i\nu} K_{\i\nu}\big(c|\eta|\big),
\]
where $K_{\i\nu}(z)$ is the modified Bessel function of the second kind (Macdonald function) of order $\i\nu$. Using the exponential decay of $K_{\i\nu}(z)$ for large $z$ and the asymptotics $K_{\i\nu}(z)\sim \Gamma(\i\nu)2^{\i\nu-1}z^{-\i\nu}+\Gamma(-\i\nu)2^{-\i\nu-1}z^{\i\nu}$ at $z=0$, and then taking the Fourier transform, we find
\[
 (c^2 + |y|^2)^{\i\nu-\D} \sim  |y|^{2\i\nu-2\D} +   \pi^{\D}\frac{\Gamma(-\i\nu)}{\Gamma(-\i\nu+\D)} c^{2\i\nu}\delta(y)
\]
at $c=0$. As in \cite[Sec.~2.8]{gelfand} we can generalize the formulae by replacing $|y|^2$ by a complex quadratic form, or by allowing $c^2$ to be complex, and obtain this way in particular
\beq\label{eq:diffasp}
 \big({\textstyle\frac{\mu}{2}}\pm \i 0 + |y|^2\big)^{\i\nu-\D} \sim  |y|^{2\i\nu-2\D} +   2^{-\i\nu}\pi^{\D}\frac{\Gamma(-\i\nu)}{\Gamma(-\i\nu+\D)} (\mu\pm\i 0)^{\i\nu}\delta(y)
\eeq
at $\mu=0$. Next, by a direct computation we find
\[
\bea
x_\ss\cdot \xi &= \frac{1}{2}\left((1+\mu)^{\12}-(1-\mu)^{\12}(\hat x_\ss\cdot \hat\xi)\right),\\
\xi_+(x)\cdot \xi &= \frac{1}{2}(1-\hat x_\ss\cdot \hat\xi)= \frac{1}{4}(  \hat x_\ss- \hat\xi)^2,\\
\eea
\]
hence
\[
(1-\mu)^{-\12}(x_\ss\cdot \xi)=\frac{1}{2}\left(\frac{1+\mu}{1-\mu}\right)^{\12}-\frac{1}{2}+\xi_+(x)\cdot \xi=\frac{\mu}{2}+\xi_+(x)\cdot \xi+O(\mu^2).
\]
The distribution $(x_\ss^\pm \cdot \xi)^{\i \nu-\D}$ has the same asymptotics at $\mu=0$ as the distribution $(\frac{\mu}{2}\pm\i 0 +\xi_+(x)\cdot \xi)^{\i\nu-\D}$. Applying \eqref{eq:diffasp} to the former yields the desired result. 
\qeds

Note that one can also deduce a $\scri_-$ version of this statement, namely, if $\xi\in \mathcal{C}_-^d$ then at $\scri_-$,
\beq\label{eq:asss12}
(x_\ss^\pm \cdot \xi)^{\i \nu-\D} \sim (\xi_-(x)\cdot\xi)^{\i\nu - \D} + (\mu(x)\mp\i 0)^{\i\nu} a(\nu) \delta(\xi_-(x),\xi).
\eeq
Turning our attention to asymptotics at $\scri_+$ again (and $\xi\in \mathcal{C}_+^d$), we observe that \eqref{eq:asss1} also gives immediately
\beq\label{eq:asss2}
(\mu\pm\i 0)^{\i\nu}(x_\ss^\pm \cdot \xi)^{-\i\nu-\D} \sim a(-\nu) \delta(\xi_+(x),\xi)+(\mu\pm\i 0)^{\i\nu}(\xi_+(x)\cdot\xi)^{-\i\nu - \D}.
\eeq
We see that the positive frequency plane waves, i.e.~those with $x_\ss^+$, have no $(\mu-\i 0)^{\i\nu}$ term, and negative frequency plane waves have no $(\mu+\i 0)^{\i\nu}$ term.

\begin{proposition}\label{prop:extwaves} The conformal extensions of the de Sitter plane waves are given by 
\beq\label{eq:extwaves}
\bea
\ex_\nu(x_\dS^\pm\cdot \xi)^{\i\nu-\D}&=(x_\ss^\pm \cdot \xi)^{\i\nu-\D},\\
\ex_\nu(x_\dS^\pm\cdot \xi)^{-\i\nu-\D}&=\e^{\pm\nu\pi}(\mu\pm\i 0)^{\i\nu}(x_\ss^\pm \cdot \xi)^{-\i\nu-\D}.
\eea
\eeq
\end{proposition}
More precisely, both sides of the identities  in \eqref{eq:extwaves}  make strict sense after smearing in $\xi$ with an arbitrary smooth function $\psi$, homogeneous of degree $-\i\nu-\D$, respectively $\i\nu-\D$. In view of Theorem \ref{thm:ext}, The proof of Proposition \ref{prop:extwaves} is a straightforward computation which consists in checking that the right hand side of each line in \eqref{eq:extwaves} is a solution for $P_\nu$ which in the de Sitter region of $\ss^d$ coincides with $f^{\i\nu-\D}(x_\dS^\pm\cdot \xi)^{\i\nu-\D}$, respectively $f^{\i\nu-\D}(x_\dS^\pm\cdot \xi)^{-\i\nu-\D}$.

Proposition \ref{prop:extwaves} together with the asymptotics \eqref{eq:asss1}, \eqref{eq:asss2} gives the asymptotic data of the conformal extensions of de Sitter plane waves, namely
\beq\label{eq:dataplane1}
\bea
\big(\varrho_{\scri_+} \ex_\nu(x_\dS^+\cdot \xi')^{\i\nu-\D}\big)(\xi)&=\begin{pmatrix} a(\nu) \delta(\xi,\xi')  \\ 0 \end{pmatrix},\\
\big(\varrho_{\scri_+} \ex_\nu(x_\dS^-\cdot \xi')^{\i\nu-\D}\big)(\xi)&=\begin{pmatrix} 0 \\ a(\nu)\delta(\xi,\xi') \end{pmatrix},\\
\eea
\eeq  
and 
\beq\label{eq:dataplane2}
\bea
\big(\varrho_{\scri_+} \ex_\nu(x_\dS^+\cdot \xi')^{-\i\nu-\D}\big)(\xi)&=\begin{pmatrix} \e^{\nu\pi} (\xi\cdot\xi')^{-\i\nu - \D} \\ 0 \end{pmatrix},\\
\big(\varrho_{\scri_+} \ex_\nu(x_\dS^-\cdot \xi')^{-\i\nu-\D}\big)(\xi)&=\begin{pmatrix} 0 \\ \e^{-\nu\pi} (\xi\cdot\xi')^{-\i\nu - \D} \end{pmatrix},\\
\eea
\eeq  
where again all identities make sense after smearing in $\xi'$ with an arbitrary smooth function of correct homogeneity. This confirms that the asymptotic data distinguish between positive and negative frequencies.


\subsection{Identities at $\scri_+$} We will need the identities 
\beq\label{eq:inv1}
\bea
(x_\dS^\pm\cdot \xi)^{-\i\nu-\D}&= \frac{\e^{\pm \nu\pi}}{a(\nu)}\int_{\scri_+} \omega(\xi')(x_\dS^\pm \cdot \xi')^{\i\nu-\D} (\xi'\cdot\xi)^{-\i\nu-\D},\\
(x_\hh\cdot \xi)^{-\i\nu-\D}&=\frac{1}{a(\nu)}\int_{\scri_+} \omega(\xi')(x_\hh \cdot \xi')^{\i\nu-\D} (\xi'\cdot\xi)^{-\i\nu-\D},
\eea
\eeq
$\xi\in \scri_+$, see \cite[Prop.~II.4]{faraut}, \cite[Sec.~4.4]{MS} and \cite[Lem.~4.1]{EM}. Both sides of the identities \eqref{eq:inv1} are understood as analytic continuation from $\Re(\i\nu)>0$ (note that the integrals in \eqref{eq:inv1} are indeed well defined if $\Re(\i\nu)>0$). Using \eqref{eq:mupm} we conclude from \eqref{eq:inv1}
\beq\label{eq:inv2}
(\mu\pm\i 0)^{\i\nu}(x_\ss^\pm\cdot \xi)^{-\i\nu-\D}=\frac{1}{a(\nu)}\int_{\scri_+} \omega(\xi')(x_\ss^\pm \cdot \xi')^{\i\nu-\D} (\xi'\cdot\xi)^{-\i\nu-\D}
\eeq
in the sense of analytic continuation from $\Re(\i\nu)>0$. 

Another useful identity is 
\beq\label{eq:delta}
\int_{\scri_+}\omega(\xi)(\xi_1\cdot\xi)^{\i\nu-\D}(\xi\cdot\xi_2)^{-\i\nu-\D} = a(\nu)a(-\nu) \delta(\xi_1,\xi_2),
\eeq
see \cite[App.~B]{HL} for a mathematically rigorous derivation and additional remarks. A direct consequence of \eqref{eq:delta} is that the inverse of the operator with kernel
\[
\cS_{\scri_+}(\xi,\xi')=\frac{1}{a(\nu)}(\xi\cdot\xi')^{\i\nu-\D}
\]
is the operator with kernel $\cS_{\scri_+}^{-1}(\xi,\xi')=\frac{1}{a(-\nu)}(\xi\cdot\xi')^{-\i\nu-\D}$. Especially in the mathematical literature, the operator $\cS_{\scri_+}$ is often called the \emph{scattering matrix} on $\hh^d$. Its special feature is that it maps $f^{\i\nu+\D}_\hh$ asymptotics of solutions to $f^{-\i\nu+\D}_\hh$ asymptotics (therefore one can think of $\cS_{\scri_+}$ as a generalized Dirichlet-to-Neumann map). The operator $\cS_{\scri_+}$ also appears in the formula for the asymptotic data $\varrho_{\scri_+}$ in terms of $f^{\i\nu+\D}_\dS$ and $f^{-\i\nu+\D}_\dS$ asymptotics. Namely, if $u$ solves the Klein-Gordon equation \eqref{eq:dflnj} and $u\sim f^{-\i\nu+\D}_\dS w^- +  f^{\i\nu+\D}_\dS w^+$, then
\[
\varrho_{\scri_+}\ex_\nu u = \frac{1}{\e^{\nu\pi}-\e^{-\nu\pi}}\begin{pmatrix}\one & -\e^{-\nu\pi} \cS_{\scri_+}^{-1} \\ -\one & \e^{\nu\pi} \cS_{\scri_+}^{-1}\end{pmatrix}\begin{pmatrix} w^+ \\ w^- \end{pmatrix},
\]
see \cite{poisson} for more general statements and their proof.

\section{Conformal extension of two-point functions}

\subsection{Formal adjoints}\label{ss:adjoints} Let $|d g_\dS|$, resp.~$|d g_\hh|$ be the volume density on $\dS^d$, resp.~$\hh^d$. The differential operator $P_\nu$ is formally self-adjoint with respect to the unique smooth density which equals $\mu^{\alpha+1}|d g_\dS|$ or $\mu^{\alpha+1}|d g_\hh|$ in the respective region (see \cite[Sec.~3.1]{resolvent}). Generally, we will use this density to define formal adjoints or Schwartz kernels on $\ss^d$. On the other hand, we use the usual volume density $|d g_\dS|$ or $|d g_\hh|$ to define formal adjoints and Schwartz kernels on $\dS^d$ or $\hh^d_\pm$.

Correspondingly, we write $e_\nu^*$ to mean the formal adjoint of $e_\nu$, defined using the two densities on respectively $\dS^d$ and $\ss^d$. With these conventions, $e_\nu^*$ maps by restriction to $\dS^d$ and multiplication by $f^{-\i\nu+\alpha+2}_\dS$.



\subsection{Bunch-Davies two-point function}

Thanks to a result by Bros and Moschella \cite{bros1,bros2}, the \emph{Bunch-Davies two-point function} can be defined as the following bi-distribution on $\dS^d$ (cf.~\cite{BD,allen} for the original definition):
\beq\label{eq:defBD}
\Lambda^+(x_\dS,y_\dS)=c_{d,\nu}\e^{-\nu\pi} \int_{\scri_+}\omega(\xi) (x_\dS^-\cdot\xi)^{\i\nu-\D}(\xi\cdot y_\dS^+)^{-\i\nu-\D},
\eeq
where 
\[
c_{d,\nu}=\frac{\Gamma(\i\nu+\D)\Gamma(-\i\nu+\D)}{2^{d+1}\pi^d}.
\]
Let us also introduce the bi-distribution
\beq\label{eq:defBDbar}
\Lambda^-(x_\dS,y_\dS)=c_{d,\nu}\e^{-\nu\pi} \int_{\scri_+} \omega(\xi)(x_\dS^+\cdot\xi)^{-\i\nu-\D}(\xi\cdot y_\dS^-)^{\i\nu-\D},
\eeq
which is simply the complex adjoint of $\Lambda^+$, i.e.
\beq\label{eq:chinv}
\Lambda^-(x_\dS,y_\dS) = \overline{\Lambda^+(x_\dS,y_\dS)}. 
\eeq
The term \emph{two-point function} means that the operator $\Lambda^+$ with Schwartz kernel $\Lambda^+(x_\dS,y_\dS)$ satisfies:
\beq\label{eq:2pt}
\begin{array}{rl}
{i})&P\Lambda^\pm=\Lambda^\pm P=0,\\[2mm]
{ii})& \Lambda^+ - \Lambda^-=\i E,\\[2mm]
{iii}) & \Lambda^\pm\geq 0 \mbox{ on } C_{\rm c}^\infty(\dS^d),
\end{array}
\eeq
where $P=\Box_\dS + \mass$ and $E$ is the causal propagator of $P$ (often also called the Pauli-Jordan commutator function). In the literature it is very common to formulate \eqref{eq:2pt} purely in terms of $\Lambda^+$ (and to use e.g.~the notation $W$ in the place of $\Lambda^+$), which after taking account \eqref{eq:chinv} gives a simpler looking set of conditions.

In view of the properties \eqref{eq:chinv} and \eqref{eq:2pt}, the Bunch-Davies two-point function $\Lambda^+$ can be used to construct (smeared, real) quantum fields 
\[
C^\infty_{\rm c}(\dS^d;\rr)\ni v\mapsto \phi(v)
\]
($\rr$-linear in $v$) in the well-known way. Equivalently, in the somewhat less familiar charged formalism, one can use $\Lambda^+$ and $\Lambda^-$ to construct charged quantum fields
\[
C^\infty_{\rm c}(\dS^d)\ni v\mapsto \psi(v)
\]
(which are $\cc$-anti-linear in $v$), and then the usual real fields are recovered using the formula $\phi(v)=\frac{1}{\sqrt{2}}(\psi(v)+\psi^*(v))$ for real valued $v$. Let us point out that in the charged formalism, only the conditions \eqref{eq:2pt} are viewed as necessary, while the identity \eqref{eq:chinv} is treated as an additional property of $\Lambda^+$, namely its invariance under a \emph{charge reversal}, which is here simply the complex conjugation.  

\subsection{Conformal extension of two-point functions} Let us denote by $r_\nu$ the conformal restriction
\[
r_\nu \tilde u \defeq f^{-\i\nu+\D}_{\dS}(\tilde u\traa{\dS^d}):\cD'(\ss^d)\to\cD'(\dS^d).
\]
The statement of Theorem \ref{thm:ext} generalizes in a straighforward way to bi-solutions. Namely, given a bi-solution $\Lambda$ of $P$, if it maps continuously $C_{\rm c}^\infty(\dS^d)$ to $C^\infty(\dS^d)$ then $e_\nu \Lambda e_\nu^*$ is the unique bi-solution of $P_\nu$ that conformally extends $\Lambda$ in the sense that $r_\nu (e_\nu \Lambda e_\nu^*) r_{\nu}^*=\Lambda$. In the special case of the Bunch-Davies two-point function it is possible to give a short-hand formula in terms of conformally extended plane waves.

\begin{theorem}\label{thm:ext2} Assume $\nu\in\rr\setminus\{0\}$. There exists a unique bi-solution $\Lambda^+_\nu$ of $P_\nu$ that conformally extends the Bunch-Davies two-point function $\Lambda^+$. Similarly, there exists a unique bi-solution $\Lambda^-_\nu$ of $P_\nu$ that conformally extends $\Lambda^-$.
Their Schwartz kernels are given by
\beq\label{eq:lambdaformula}
\Lambda^\pm_\nu(x_\ss,y_\ss)= c_{d,\nu}\e^{\mp\nu\pi} \int_{\scri_+} \omega(\xi)(x_\ss^\mp\cdot\xi)^{\i\nu-\D}(\xi\cdot y_\ss^\pm)^{-\i\nu-\D} .
\eeq
\end{theorem}
\proof The formula for $\Lambda^+_\nu=e_\nu \Lambda^+ e_\nu^*$ follows directly from \eqref{eq:defBD} and Proposition \ref{prop:extwaves}. 
By \eqref{eq:defBD} and Proposition \ref{prop:extwaves} the Schwartz kernel of $\Lambda^-_\nu=e_\nu \Lambda^- e_\nu^*$ is
\[
\Lambda^-_\nu(x_\ss,y_\ss)=c_{d,\nu}\e^{\nu\pi} \int_{\scri_+}\omega(\xi)(\mu(x)+\i0)^{\i\nu} (x_\ss^+\cdot\xi)^{-\i\nu-\D}(\xi\cdot y_\ss^-)^{\i\nu-\D}(\mu(y)-\i0)^{-\i\nu}.
\]
Using \eqref{eq:inv2} twice we get that $\Lambda^-_\nu(x_\ss,y_\ss)$ equals
\[
\frac{c_{d,\nu}\e^{\nu\pi}}{a(\nu)a(-\nu)} \iiint_{\scri_+^3}\omega(\xi)\omega(\xi_1)\omega(\xi_2) (x_\ss^+\cdot\xi_1)^{\i\nu-\D}(\xi_1\cdot\xi)^{-\i\nu-\D}(\xi\cdot\xi_2)^{\i\nu-\D}(\xi_2\cdot y_\ss^-)^{-\i\nu-\D}.
\]
By \eqref{eq:delta}, integrating in $\xi$ gives
\[
\Lambda^-_\nu(x_\ss,y_\ss)=c_{d,\nu}\e^{\nu\pi}\iint_{\scri_+^2}\omega(\xi_1)\omega(\xi_2) (x_\ss^+\cdot\xi_1)^{\i\nu-\D}\delta(\xi_1,\xi_2)(\xi_2\cdot y_\ss^-)^{-\i\nu-\D},
\]
and by integrating in either $\xi_1$ or $\xi_2$ we obtain the desired formula. \qeds

Let $E_\nu$ be the conformal extension of the causal propagator $E$, i.e. $E_\nu = e_\nu E e_\nu^*$. The conformally extended two-point functions (or more precisely, the associated operators, defined using the density described in Subsect.~\ref{ss:adjoints}) satisfy
\beq
\begin{array}{rl}
{i})&P_\nu\Lambda_{\nu}^\pm=\Lambda_{\nu}^\pm P_\nu=0,\\[2mm]
{ii})& \Lambda_\nu^+ - \Lambda_\nu^-=\i E_\nu,\\[2mm]
{iii}) & \Lambda_\nu^\pm\geq 0.
\end{array}
\eeq
Note that in contrast to the original two-point functions $\Lambda^+$ and $\Lambda^-$, in the case of $\Lambda^+_\nu$ and $\Lambda^-_\nu$ it is no longer true that the Schwartz kernel of one is the complex conjugate of the other. Instead, $\Lambda^+_\nu$ is invariant under the charge reversal $\kappa u \defeq e_\nu \overline{r_{\nu}u}$ (here meant as an anti-linear involution, defined for convenience on solutions of $P_\nu$; see \cite{VW} for a precise description of the relationships between the various symplectic spaces)  in the sense that
\[
\kappa_\nu (\Lambda^+_\nu)\kappa_\nu^*=\Lambda^-_\nu.
\] 

On the side note, we remark that the conformal restrictions of $\Lambda^+_\nu$, $\Lambda^-_\nu$ and $E_\nu$ to $\hh^d_+$ are \emph{all} proportional to the positive half-line spectral projector of $-\Delta_\hh-\mmass$, see \cite{VW}, alternatively this can be checked using \eqref{relplw} and the formulae in \cite{MS}.

\subsection{Symplectic form at $\scri_+$ and Poisson operators} 

Using \eqref{eq:lambdaformula} and \eqref{eq:dataplane1}, we can compute the following Schwartz kernels:
\[
\bea
(\varrho_{\scri_+}\circ \Lambda^+_\nu)(\xi,y_{\ss})&=c_{d,\nu}a(\nu)\big(0 ,\e^{-\nu\pi}(\xi\cdot y_\ss^+)^{-\i\nu-\D}\big)^{\rm t},\\
(\varrho_{\scri_+}\circ \Lambda^-_\nu)(\xi,y_{\ss})&=c_{d,\nu}a(\nu)\big(\e^{\nu\pi}(\xi\cdot y_\ss^-)^{-\i\nu-\D},0\big)^{\rm t}.
\eea
\]
Using this it is straightforward to compute the Schwartz kernel of $\varrho_{\scri_+}\circ E_\nu$ and $(\varrho_{\scri_+}\circ E_\nu)^*$ and to check the identity
\beq\label{eq:sympl}
(\varrho_{\scri_+}\circ E_\nu)^*\circ q \circ (\varrho_{\scri_+}\circ E_\nu) = \i E_\nu,
\eeq
where 
\[
q=\frac{1}{c_{d,\nu}a(\nu)a(-\nu)}\begin{pmatrix}-\e^{-\nu\pi} & 0 \\ 0 & \e^{\nu\pi}\end{pmatrix}.
\]
From \eqref{eq:sympl} we conclude that $\i^{-1} E_\nu^*\varrho_{\scri_+}^* q$ is a left inverse of $\varrho_{\scri_+}$ on the range of $E_{\nu}$, and thus (by \cite[Prop.~2.2]{VW}), on solutions of $P_\nu u=0$. The computation of the Schwartz kernel of that left inverse gives the following short-hand formula.

\begin{proposition}\label{prop:recover} For any $v\in\cf(\scri_+)^2$, the unique solution $u$ of $P_\nu u=0$ with data $\varrho_{\scri_+} u = v$ is given by $\cU_{\scri_+} v$, where $\cU_{\scri_+}=\i^{-1} E_\nu^*\varrho_{\scri_+}^* q$ has Schwartz kernel
\[
\cU_{\scri_+}(x_{\ss},\xi)=\frac{1}{a(\nu)}\big( (x_\ss^+\cdot \xi)^{\i\nu-\D}, (x_\ss^-\cdot \xi)^{\i\nu-\D} \big).
\]
\end{proposition}

Statement (1) in Theorem \ref{thm:main} follows now in a straightforward way. Statement (2) in the same theorem (i.e., the fact that the construction in \cite{VW} yields the Bunch-Davies state) simply follows from the observation that $\Lambda_\nu^\pm$ is proportional to $(\varrho_{\scri_+}\circ E_\nu)^*\circ \pi^\pm \circ (\varrho_{\scri_+}\circ E_\nu)$, where $\pi^\pm$ projects to the respective component.

A further consequence is that Vasy's Feynman propagator, defined in \cite[Sec.~4]{positive} in terms of suitable boundary conditions at $\scri_+$ and $\scri_-$, coincides with the Feynman propagator canonically associated with the Bunch-Davies state. This follows directly by comparing their asymptotic properties, as Vasy's boundary conditions consist in requiring that both $\varrho_{\scri_+}$ data and $\varrho_{\scri_-}$ data have only one component. This can be thought of as the analogue of the fact that on Minkowski space, the Feynman inverse discussed in \cite{GHV} or \cite{GWfeynman,GWfeynman2,DS} coincides with the well-familiar Feynman propagator canonically associated with the vacuum state.

\subsection*{Acknowledgments} Support from the grant ANR-16-CE40-0012-01 is gratefully acknowledged.


\begin{thebibliography}{plain}

\bibitem{allen} B.~Allen, {\em Vacuum States in de Sitter Space}, Phys. Rev. D 32, 3136 (1985).

\bibitem{ashtekar} A.~Ashtekar, M.~Streubel, {\em Symplectic geometry of radiative modes and conserved quantities at null infinity}, Proc. R. Lond. A 376, (1981), 585.

\bibitem{BB} J.C.A.~Barata, M.~Brum, {\em Wavepackets on de Sitter spacetime}, preprint \texttt{arXiv:1708.00538} (2017).

\bibitem{BJM} J.C.A.~Barata, C.~J\"akel, J.~Mund {\em Interacting quantum fields on de Sitter Space}, preprint \texttt{arXiv:1607.02265} (2017).











\bibitem{bros1} J.~Bros, U.~Moschella, {\em Two point functions and quantum fields in de
Sitter universe}, Rev. Math. Phys. 8, 327 (1996).

\bibitem{bros2} J.~Bros, U.~Moschella, J.~P.~Gazeau, {\em Quantum field theory in the de
Sitter universe}, Phys. Rev. Lett. 73, 1746 (1994).

\bibitem{bros3} J.~Bros, H.~Epstein, U.~Moschella, {\em Particle decays and stability on the de Sitter universe}, Ann. Henri Poincar\'e 11, (2010), 611.

\bibitem{BD} T.S.~Bunch, P.C.W.~Davies: Proc. R.Soc. Lond. A 360, 117 (1978).




\bibitem{DMP} C.~Dappiaggi, V.~Moretti and N.~Pinamonti, {\em Rigorous steps towards holography in asymptotically flat spacetimes},
Rev. Math. Phys. 18, 349 (2006).
 
\bibitem{DMP1} C.~Dappiaggi, V.~Moretti and N.~Pinamonti, {\em Distinguished quantum states in a class
of cosmological spacetimes and their Hadamard property}, J. Math. Phys. 50 (2009) 062304.

\bibitem{DMP2} C.~Dappiaggi, V.~Moretti and N.~Pinamonti, {\em Rigorous construction and Hadamard property of the Unruh state in Schwarzschild spacetime}, Adv. Theor. Math. Phys. 15 (2011) 355.


\bibitem{DS} J.~Derezi\'nski, D.~Siemssen, {\em Feynman propagators on static spacetimes}, Rev. Math. Phys. 30, 03, (2018) 1850006.




\bibitem{EM} H.~Epstein, U.~Moschella, {\em de Sitter tachyons and related topics}, Commun. Math. Phys. 336, (1) (2015),
381--430.

\bibitem{faraut}  J.~Faraut, {\em Noyaux sph\'eriques sur un hyperboloide \`a une nappe}, Lecture Notes in Math. 497, Springer
Verlag, (1975).


\bibitem{friedlander} F.~G.~Friedlander, {\em Radiation fields and hyperbolic scattering theory}, Math. Proc. Camb. Phil. Soc. 88 (1980), 483--515.


                                 
																																																																				\bibitem{gelfand} I.~M.~Gel'fand and G.~E.~Shilov, {\em Generalized Functions, Volume 1}, Academic Press, 1994.
																																																							
																																																																																																																														
\bibitem{characteristic} C.~G\'erard, M.~Wrochna, \textsl{Construction of Hadamard states by characteristic Cauchy problem}, Anal. PDE 9 (1), 111-149 (2016).

\bibitem{inout}  C.~G\'erard, M.~Wrochna, \textsl{Hadamard property of the \emph{in} and \emph{out} states for Klein-Gordon fields on asymptotically static spacetimes},  Ann. Henri Poincar\'e 18 (8), (2017), 2715--2756.

\bibitem{GWfeynman}  C.~G\'erard, M.~Wrochna, \textsl{The massive Feynman propagator on asymptotically Minkowski spacetimes I}, to appear in Amer. J. Math., \texttt{arXiv:1609.00192} (2016).

\bibitem{GWfeynman2}  C.~G\'erard, M.~Wrochna, \textsl{The massive Feynman propagator on asymptotically Minkowski spacetimes II}, preprint \texttt{arXiv:1806.05076} (2018).




\bibitem{GHV} J.~Gell-Redman, N.~Haber, A.~Vasy, \textsl{The Feynman propagator on perturbations of Minkowski space}, Comm. Math. Phys. 342 (1), (2016), 333--384.






\bibitem{HN} D.~H\"afner, J.-P.~Nicolas, {\em The characteristic Cauchy problem for Dirac fields on curved backgrounds}, J. of Hyperbolic Differ. Equ. 8 (2011), 437--483.



\bibitem{HL} S.~Hollands, G.~Lechner, {\em SO(d,1)-invariant Yang-Baxter operators and the ${\rm dS}/{\rm CFT}$ correspondence}, Commun. Math. Phys., DOI 10.1007/s00220-017-2942-6 (2017).

\bibitem{HW} S.~Hollands, R.~M.~Wald, {\em Quantum fields in curved spacetime}, in: General Relativity and Gravitation: A Centennial Perspective, Cambridge University Press (2015).










\bibitem{mason} L.~J.~Mason, J.-P.~Nicolas, {\em Conformal scattering and the Goursat problem}, J. Hyperbolic Differ. Equ. 1 (2004), 2, 197--233.



\bibitem{melrose2} R.~Melrose, \textsl{Spectral and scattering theory for the Laplacian on asymptotically Euclidean spaces}, Lecture Notes in Pure and Appl. Math., vol. 161, Dekker, New York (1994), 85--130.


\bibitem{Mo1} V.~Moretti, {\em  Uniqueness theorem for BMS-invariant states of scalar QFT on the null boundary of asymptotically flat spacetimes and bulk-boundary observable algebra correspondence}, Comm. Math. Phys. {268} (2006), 727-756.

\bibitem{Mo2} V.~Moretti, {\em  Quantum out-states holographically induced by asymptotic flatness: invariance under space-time symmetries, energy positivity and Hadamard property}, Comm. Math. Phys. {279} (2008), 31-75.

\bibitem{MS} U.~Moschella, R.~Schaeffer, {\em Quantum Theory on Lobatchevski Spaces}, Class. Quant. Grav., 24:3571--3602, (2007).

\bibitem{nicolas} J.-P.~Nicolas, {\em Conformal scattering on the Schwarzschild metric}, Ann. Inst. Fourier (Grenoble), 66, no. 3, (2016), 1175--1216.

\bibitem{penrose64} R.~Penrose, {\em Conformal treatment of infinity}, Relativit\'e, Groupes et Topologie (Lectures,
Les Houches, 1963 Summer School of Theoret. Phys., Univ. Grenoble), (1964), 563--584.

\bibitem{penrose65} R.~Penrose, {\em Zero rest-mass fields including gravitation: asymptotic behaviour}, Proc. Roy. Soc. London A284 (1965), 159--203.

\bibitem{queva} J.~Qu\'eva, {\em Sur quelques probl\`emes de quantification : en espace-temps de de Sitter et par \'etats coh\'erents}, Ph.D. thesis, Universit\'e Paris Diderot -- Paris 7, \texttt{http://tel.archives-ouvertes.fr/tel-00503186/fr/}, (2009). 







\bibitem{resolvent} A.~Vasy, \textsl{Microlocal analysis of asymptotically hyperbolic spaces and high energy resolvent estimates}, MSRI Publications, vol. 60, Cambridge University Press, (2012).


\bibitem{positive} A.~Vasy, \textsl{On the positivity of propagator differences},  Ann. Henri Poincar\'e, 18, (2017), 983--1007. 

\bibitem{kerrds} A.~Vasy, \textsl{Microlocal analysis of asymptotically hyperbolic and Kerr-de Sitter spaces, (With an appendix by S. Dyatlov)}, Inventiones Math., 194:381–513, (2013), 194.2: 381-513.

 

\bibitem{poisson} A.~Vasy, \textsl{Resolvents, Poisson operators and scattering matrices on asymptotically hyperbolic
and de Sitter spaces}, J. Spect. Theory, vol. 4 (4), (2014), 643--673.



 
\bibitem{VW} A.~Vasy, M.~Wrochna, {\em  Quantum fields from global propagators on asymptotically Minkowski and extended de Sitter spacetimes}, Ann. Henri Poincar\'e, 19 (5), (2018), 1529--1586.



\end{thebibliography}
\end{document}